\documentstyle[12pt,thmsa,sw20lart]{article}
\begin{document}

\title{Neutrinos: recent developments and origin of neutrino mass matrix\thanks{%
Based on the talk delivered at First International Conference on Modern
Trends in Physics Research (MTPR-04), April 04-09, 2004, Egypt; to appear in
the proceedings.}}
\author{Riazuddin \\
National Center for Physics, Quaid-i-Azam University,\\
Islamabad 45320, Pakistan.}
\date{May 2004, NCP-QAU/2004-005}
\maketitle

\begin{abstract}
Certainly one of the most exciting areas of research at present is neutrino
physics. The neutrinos are fantastically numerous in the universe and as
such they have bearing on our understanding of the universe. Therefore, we
must understand the neutrinos, particularly their mass. There is compelling
evidence from solar and atmospheric neutrinos and those from reactors for
neutrino oscillations implying that neutrinos mix and have nonzero mass but
without pinning down their absolute mass. This is reviewed. The implications
of neutrino oscillations and mass squared splitting between neutrinos of
different flavor on pattern of neutrino mass matrix is discussed. In
particular, a neutrino mass matrix, which shows approximate flavor symmetry
where the neutrino mass differences arise from flavor violation in
off-diagonal Yukawa couplings is elaborated on. The implications in double
beta decay are also discussed.
\end{abstract}

\newpage\ 

\section{Introduction}

Certainly one of the most exciting areas of research at present is neutrino
physics. Neutrinos are fantastically numerous in the universe and as such to
understand the universe we must understand neutrinos. It is fair to say that
the results of the last decade on neutrinos from the sun, from the
atmospheric interaction of cosmic rays, and from reactors provide a
compelling evidence that the neutrinos have nonzero mass and mix.

In 1930's protons, neutrons and electrons were considered as elementry
particles. Such a picture was confronted with two fundamental problems:
conservation of energy and angular momentum (A.M.) in $\beta $-decay 
\[
\,n\rightarrow p+e^{-} 
\]
This is because experimentally seen continuous $\beta $ spectrum can not be
explained for 2-body final state if energy is conserved, since in that case $%
E_e$ would have an unique energy. Further the final state would necessarily
have integral A.M. while initial state has half integeral A.M.

To solve these problems Pauli assumed that there exists a new electrically
neutral elementry particle, with spin $1/2$, mass less than electron mass
and an interaction much weaker than photon interaction. Thus 
\[
\,n\rightarrow p+e^{-}+\bar{\nu}_e 
\]
leading to continuous $\beta $ spectrum and conservation of A.M. This was
the first particle postulated by a theoretician.

Direct observation of $\bar{\nu}_e$ was made much later in 1950's, when high
flux fission reactors as source of neutrons become available. $\bar{\nu}_e$%
's, electron-type antineutrinos are produced in the decay of pile neutrons
in a fission reactor. These can be captured in hydrogen giving the reaction 
\[
\bar{\nu}_e+p\rightarrow n+e^{-} 
\]
whose cross-section was measured by Reines and Cowan: 
\[
\sigma _{\exp }=\left( 11\pm 2.5\right) \times 10^{-44}{ cm}^2 
\]
to be compared with the theoretical value 
\[
\sigma _{th}=\left( 11\pm 1.6\right) \times 10^{-44}{ cm}^2 
\]
Note the extreme smallness of the cross-section (nuclear cross sections are
of order $10^{-24}$cm$^2$). It is a reflection of the fact that neutrino has
only weak interaction. It is remarkable that neutrinos which have almost no
interaction with matter have contributed to some of the most important
discoveries in physics given below:

\begin{itemize}
\item  1950's \thinspace \thinspace \thinspace \thinspace \thinspace
\thinspace \thinspace \thinspace \thinspace \thinspace \thinspace \thinspace 
$\bar{\nu}_e$: electron type anti-neutrino discovered in experiments of
Reines and Cowan (1995 Nobel Prize).

\item  1956: \thinspace \thinspace \thinspace \thinspace \thinspace
\thinspace \thinspace \thinspace \thinspace \thinspace \thinspace \thinspace
\thinspace \thinspace \thinspace \thinspace \thinspace \thinspace \thinspace
\thinspace \thinspace \thinspace \thinspace \thinspace \thinspace \thinspace
Parity non-conservation in $\beta $ decays was discovered (Wu et al.) after
its conservation in weak interaction was questioned by Lee and Yang (1957
Nobel Prize).

\item  1957: \thinspace \thinspace \thinspace \thinspace \thinspace
\thinspace \thinspace \thinspace \thinspace \thinspace \thinspace \thinspace
\thinspace \thinspace \thinspace \thinspace \thinspace \thinspace \thinspace
\thinspace \thinspace \thinspace \thinspace \thinspace \thinspace \thinspace
It was proved that neutrino (antineutrino) is left handed (right handed)
particle (Goldharber etal), after Salam, Landau and Lee and Yang proposed
the 2-component neutrino theory.

\item  1962: \thinspace \thinspace \thinspace \thinspace \thinspace
\thinspace \thinspace \thinspace \thinspace \thinspace \thinspace \thinspace
\thinspace \thinspace \thinspace \thinspace \thinspace \thinspace \thinspace
\thinspace \thinspace \thinspace $\nu _\mu :$\thinspace The muon neutrino
was discovered (Lederman, Steinberger, Schwartz et al.)\thinspace (1988
Nobel Prize).

\item  1970:\thinspace \thinspace \thinspace \thinspace \thinspace
\thinspace \thinspace \thinspace \thinspace \thinspace \thinspace \thinspace
\thinspace \thinspace \thinspace \thinspace \thinspace \thinspace \thinspace
\thinspace \thinspace \thinspace \thinspace \thinspace \thinspace \thinspace
Solar neutrinos were detected in pioneering experiments by R. Davis.

\item  1973: \thinspace \thinspace \thinspace \thinspace \thinspace
\thinspace \thinspace \thinspace \thinspace \thinspace \thinspace \thinspace
\thinspace \thinspace \thinspace \thinspace \thinspace \thinspace \thinspace
\thinspace \thinspace \thinspace \thinspace \thinspace \thinspace \thinspace
A new class of weak interactions (neutral currents) was discovered in a
neutrino experiments by Garamelle Collaboration at CERN, as predicted by
electroweak unification (1979 Nobel Prize: Glashow, Salam, Weinberg).

\item  1980's: \thinspace \thinspace \thinspace \thinspace \thinspace
\thinspace \thinspace \thinspace \thinspace \thinspace \thinspace \thinspace
\thinspace \thinspace \thinspace \thinspace \thinspace \thinspace \thinspace
\thinspace \thinspace In experiments on neutrino beams at CERN and at
Fermilab, the quark structure of nucleon was established and investigated J.
I. Friedman, H. W. Kendall, R. E. Taylor (1990 Nobel Prize).

\item  1983: \thinspace \thinspace \thinspace \thinspace \thinspace
\thinspace \thinspace \thinspace \thinspace \thinspace \thinspace \thinspace
\thinspace \thinspace \thinspace \thinspace \thinspace \thinspace \thinspace
\thinspace \thinspace \thinspace \thinspace \thinspace \thinspace \thinspace
The intermediate $W$ and $Z$ bosons were discovered at CERN at masses
predicted by electro-weak unification (1984 Nobel Prize: C. Rubbia and Simon
Van Der Meer).

\item  1987: \thinspace \thinspace \thinspace \thinspace \thinspace
\thinspace \thinspace \thinspace \thinspace \thinspace \thinspace \thinspace
\thinspace \thinspace \thinspace \thinspace \thinspace \thinspace \thinspace
\thinspace \thinspace \thinspace \thinspace \thinspace \thinspace \thinspace
Neutrinos from supernova 1987 A were detected \thinspace (Kamiokanda, IMB,
Bakson).

\item  1990's:\thinspace \thinspace \thinspace \thinspace \thinspace
\thinspace \thinspace \thinspace \thinspace \thinspace \thinspace \thinspace
\thinspace \thinspace \thinspace \thinspace \thinspace \thinspace \thinspace
\thinspace \thinspace \thinspace \thinspace It was found in LEP experiments
that only three types of light flavor neutrinos exist in nature: $\nu _e$,
\thinspace $\nu _\mu $, $\nu _\tau $.

\item  2000:\thinspace \thinspace \thinspace \thinspace \thinspace
\thinspace \thinspace \thinspace \thinspace \thinspace \thinspace \thinspace
\thinspace \thinspace \thinspace $\nu _\tau :$ Direct observation of nu tau
was made (Fermi Lab's Tevatron).

\item  2001:\thinspace \thinspace \thinspace \thinspace \thinspace
\thinspace \thinspace \thinspace \thinspace \thinspace \thinspace \thinspace
\thinspace \thinspace \thinspace \thinspace \thinspace \thinspace \thinspace
\thinspace \thinspace \thinspace Solar neutrino Oscillations are
established; solar model is verified: Super Kamiokande (SK), Sadbury
Neutrino Observatory (SNO), (2002 Nobel Prize: Davis and Koshiba).
\end{itemize}

This is an impressive list of discoveries.

\section{Neutrino Mass}

Neutrino occurs in one helicity state (Left handed). This togather with
lepton number $L$ conservation implies $m_\nu =0$. However there is no deep
reason that it should be so. There is no local gauge symmetry and no
massless gauge boson coupled to lepton number $L$, which therefore is
expected to be violated. Thus one may expect a finite mass for neutrino.
Moreover, all other known fermions, quarks and charged leptons, are massive.
But the intriguing question is: why $m(\nu _e)\ll m(e)$, which needs to be
understood, even though we do not understand why e.g. electron mass is what
it is and why muons and tauons are heavier than electron see Fig. 1 [1].
This is the so called flavor problem which has so far eluded us. Neutrino
mass has added importance for two other reasons:

\begin{itemize}
\item  The interesting phenomena of neutrino oscillations is possible if one
or more of neutrinos have non vanishing mass.

\item  Non-vanishing of neutrino mass has important implications in
Astrophysics and Cosmology. It is a candidate for hot dark matter.
\end{itemize}

\subsection{Astrophysical Constraint on Neutrino mass}

The total mass-energy of the universe is composed of several constituents,
each of which is characterized by its energy density, which is expressed in
terms of critical density 
\[
\rho _{0i}\equiv \Omega _{oi}\rho _{c0} 
\]
Critical density is the minimum density required for the expansion of the
Universe to be turned around by the gravitational attraction of all the
matter in it and is defined as 
\begin{equation}
\rho _{c0}=\frac{3H_0^2}{8\pi G_N}  \label{e01}
\end{equation}
where $H_0$ is the Hubble constant and $G_N$ is the Newton's gravitational
constant Using the present value of $H_0$ (Hubble constant), namely $%
H_0=100h_0$ km s$^{-1}$Mpc$^{-1}$, Mpc$=${}$3\times 10^{19}$ km so that 
\[
H_0=h_0\left( 1\times 10^{10}{ yr}\right) ^{-1} 
\]
where 
\begin{equation}
h_0=0.72\pm 0.05.  \label{e02}
\end{equation}
This gives 
\begin{eqnarray}
\rho _{c0} &=&1.879h_0^2\times 10^{-29}{ g cm}^{-3}  \nonumber \\
&=&1.054h_0^2\times 10^4{ eV cm}^{-3}  \label{e03}
\end{eqnarray}

What is the neutrinos contribution to hot dark matter (since relic light $%
\nu $'s had relativistic velocity) ? Neutrinos are fantastically numerous 
\begin{eqnarray*}
n_\nu &=&\frac 3{10}n_\gamma =112\,{cm}^{-3} \\
n_\gamma &=&400\,{cm}^{-3}
\end{eqnarray*}
so if they have even a tiny mass, they can outweigh all the stars and
galaxies in the universe. The neutrinos contribution to energy density is 
\begin{eqnarray*}
\rho _{\nu _0}=\left( 112\right) \left( \sum_im_{\nu i}\,\,{eV}\right) 
{eV cm}^{-3}
\end{eqnarray*}
so that 
\begin{eqnarray}
\Omega _\nu ^{{HDM}} &=&\frac{\rho _{\nu _0}}{\rho _{c0}}  \nonumber \\
&=&\frac{112}{1.05}h_0^2\sum_im_{\nu i}\,\,\left( {eV}\right)  \nonumber
\\
&=&\frac{\sum_im_{\nu i}\,\,\left( {eV}\right) }{93.8h_0^2}  \label{e04}
\end{eqnarray}
Unfortunately there is no direct particle physics evidence on $\sum_im_{\nu
i}$. We shall come back to this question later. Here we simply note that $%
\rho _{\nu _0}\leq \rho _{c0}$, implies that 
\begin{equation}
\sum_im_{\nu i}\leq 93.8h_0^2\,\,\left( {eV}\right) =49\,{eV}
\label{e05}
\end{equation}
This is the astrophysical constraints on light neutrino masses.

\subsection{Double $\beta $-Decay}

The double $\beta $-decay is another way to look for a finite mass of
neutrino. Two kinds of double $\beta $-decay can be considered: 
\begin{eqnarray*}
(2\nu ) &&(A,\,Z)\to (A,\,Z+2)+2e^{-}+2\bar{\nu}_{e} \\
(0\nu ) &&\hspace{1.3cm}\to (A,\,Z+2)+2e^{-}.
\end{eqnarray*}

Usually the neutrinos are assumed to be Dirac particles: neutrino $\nu $ and
antineutrino $\bar{\nu}\equiv \nu ^c\,\,$are distinct.In Majorana picture,
they are identical. Thus 
\begin{eqnarray*}
n &\to &p+e^{-}+\bar{\nu}_L\equiv p+e^{-}+\nu _L \\
\nu _L+n &\to &p+e^{-},
\end{eqnarray*}
so that neutrinoless double beta decay 
\[
(2n)\to (2p)+2e^{-} 
\]
is possible. The important physics issues in $(0\nu )$ double $\beta $-decay
are:

\begin{enumerate}
\item[(i)]  Lepton number must not be conserved, which is possible if
neutrinos are Majorana particles: $\nu \equiv \bar{\nu}$

\item[(ii)]  Helicity of the neutrino cannot be exactly $-1$, this can be
satisfied if $m_\nu \neq 0$.
\end{enumerate}

Thus $(0\nu )\beta \beta $--decay is especially interesting: 
\begin{equation}
T_{1/2}\propto Q^{-5}<m_\nu >^{-2},  \label{e06}
\end{equation}
where decay $Q$ value $\approx T_{e1}+T_{e2}$. Here 
\begin{equation}
<m_\nu >=\sum_i\lambda _i\left| U_{ei}\right| ^2m_{\nu _i}  \label{e07}
\end{equation}
where $\lambda _i$ is a possible sign since Majorana neutrinos are CP
eigenstates; as shown the expectation value is weighted by neutrino's
electron couplings. There is direct evidence of $\left( 2\nu \right) \beta
\beta $ decay 
\begin{eqnarray}
\left( 2\nu \right) \beta \beta \,\,\,\,\,\,\,\,\,\,\,^{82}Se &\rightarrow
&\;^{82}Kr  \nonumber \\
T_{1/2} &=&\left( 1.1_{-0.3}^{+0.8}\right) \times 10^{20}{ yrs}
\label{e08}
\end{eqnarray}
For neutrinoless double $\beta $ decay 
\begin{eqnarray}
^{76}Ge &\rightarrow &\,^{76}Se+2e^{-}  \nonumber \\
T_{1/2} &\geq &1.9\times 10^{25}{ yrs}  \label{e09}
\end{eqnarray}
One recent result [2] has claimed the evidence for this decay with the best
value $T_{1/2}=1.5\times 10^{25}$ yrs. This analysis claims $\left\langle
m_\nu \right\rangle =\left( 0.39_{-0.28}^{+0.17}\right) $ eV which has been
commented upon [3]. If the above finding were to be confirmed, it would be
the first indication of lepton number violation in nature and that Majorana
neutrino can exist in nature. We shall come to other implications of above
value of $\left\langle m_\nu \right\rangle $ later.

\subsection{Cosmological Constraints}

We can see the universe 300,000 years after Big Bang by studying the cosmic
microwave background radiation (CMB), which is a direct relic of the
universe when it became transparent to electromagnetic radiation.
Fluctuations in the CMB radiation (at the level of a few parts in 10$^5$)
have been detected with angular resolutions from 7$^o$ to a few arc minutes
in the sky [4]. These indicate the first clumping of matter particles into
cosmic structures, which is resisted by the repulsive pressure of photons.
The net result was gravity driven acoustic--like oscillations. These
oscillations left their signature in the anisotropy of the CMB. Since the
amplitude and position of the primary and secondary peaks are directly
determined by the sound speed (and hence the equation of state) and by the
geometary and expansion of the universe, they can be used as powerful test
of the density of baryons and dark matter (DM) and other cosmological
parameters.

Recent measurements of the fluctuations by an orbiting observatory called
the Wilkinson Microwave Anisotropy Probe (WMAP) and their analysis have
settled a number of issues about the universe, its age, its expansion rate
and its composition. The results are summarized below [4]. 
\begin{eqnarray}
{Age of Universe} &=&13.4\pm 0.3{ billion years.}  \nonumber \\
\Omega &=&\rho /\rho _c=1.02\pm 0.02  \nonumber \\
\rho _{DM} &=&\left( 2.25\pm 0.38\right) \times 10^{-27}{ Kg/m}^3 
\nonumber \\
\Omega _{DM} &=&0.23\pm 0.05  \nonumber \\
\Omega _b &=&0.046\pm 0.005  \nonumber \\
\Omega _\nu &<&0.015,\,\,m_\nu <0.23\,{eV}  \label{e10}
\end{eqnarray}
Note that visible baryon density is only about 4.6 percent. The situation is
summarized in Figure 2 [5]. On the composition of the universe there is
dramatic observation that the fraction of cosmic mass-energy residing in
ordinary matter is only about 4 \%. Around 23 \% of the universe is made up
of another substance, called dark matter, proposed 25 years ago when it
became clear that all the galaxies behaved as if they were more massive than
they seemed to be. The remaining 72 \% is a new discovery, called dark
energy, that work against gravity on large scales implying that the
expansion of the universe is speeding up, rather than decelerating. In
essence what we have learned about the universe is largely restricted to 4
\%. The nature of 96 \% is essentially unknown. One thing is certain that we
have to go beyond the ordinary matter and radiation we already know. For the
dark matter we have a real chance of learning within the next 5 to 10 years
when we might discover a new type of matter at CERN, Geneva where world's
largest accelerator is being developed. Such a matter is predicted by a new
symmetry in particle physics, called supersymmetry. For dark energy, we have
to wait unless or until there is a unified theory of space-time, trying to
bring gravity within the same framework as other interactions.

\section{Origion of Neutrino Masses}

The minimal standard model involves 3 chiral neutrino states, but it does
not admit renormalizabile interactions that can generate neutrino masses. If
there is no $SU_L(2)\times U_Y(1)$-singlet fermion in nature, then neutrino
masses are necessarily Majorana 
\begin{eqnarray}
{\cal L}^{{mass}} &=&\frac 12m\psi ^TC^{-1}\psi +h.c.  \label{e11} \\
\Delta L &=&2  \nonumber
\end{eqnarray}
However, even if such a field exist, its mass is naturally much greater than
the weak scale in which case light neutrinos are Majorana fermions as we
shall see. In the SM, Majorana neutrino masses are forbidden by a global
Baryon--Lepton $\left( B-L\right) $ symmetry but there is no reason to
expect that this symmetry is fundamental. If one allows right-handed
neutrinos $\nu _R$ which are $SU_L(2)\times U_Y(1)$ singlets, then one can
write Yukawa interactions: 
\begin{equation}
{\cal L}_Y=\bar{\ell}_{Li}\phi h_{ij}e_{Rj}+\bar{\ell}_{Li}\tilde{\phi}%
h_{ij}\nu _{Rj}-\frac 12\bar{\nu}_R^cM\nu _R+h.c.  \label{e12}
\end{equation}
where the SM places the left-handed components of charged leptons and
associated neutrinos into SU$_L$(2) doublets $\ell _L$. $\phi $ is the usual
Higgs doublet under SU$_L$(2). The lepton number violation is induced by the
third term, which is allowed by the gauge symmetry. $M$ is the Majorana mass
matrix while $h$ are Yukawa couplings. After spontaneous symmetry breaking
the vacuum expectation value of the Higgs field $\left\langle \phi
\right\rangle \equiv v=175$ GeV generates the Dirac mass term $%
(m_D)_{ij}=h_{ij}v$ and 6$\times 6$ neutrino mass matrix 
\begin{equation}
M_\nu =\left( 
\begin{tabular}{ll}
0 & $m_D$ \\ 
$m_D$ & $M$%
\end{tabular}
\right)  \label{e13}
\end{equation}
After diagonalization $M_\nu $ has 6 mass eigenstates $\nu _k$ that
represent Majorana neutrinos $(\nu _k=\bar{\nu}_k)$. One can consider some
useful limits:

\begin{itemize}
\item  Dirac: $M\rightarrow 0$: there are 6 Majorana neutrinos that merge to
form 3 massive Dirac neutrinos

\item  Majorana: $m_D\rightarrow 0$

\item  Seasaw $m_D<<M$: there are three light active Majorana neutrinos
\end{itemize}

In the seesaw limit, the diagonalization of $M_\nu $ gives 
\begin{equation}
M_\nu =-m_DM^{-1}m_D  \label{e14}
\end{equation}
This also yields light and heavy neutrino mass eigen states 
\begin{eqnarray}
\nu &=&V_\nu ^T\nu _L+\nu _L^cV_\nu ^{*}  \nonumber \\
N &\approx &\nu _R+\nu _R^c  \nonumber \\
m_{Ni} &=&M_i  \label{e15}
\end{eqnarray}
where $V_\nu $ is the neutrino mixing matrix. Thus 
\begin{equation}
m_{\nu \ell }\sim \frac{m_D^2}M\approx \frac{v^2}M\ll m_\ell ,  \label{e16}
\end{equation}
by requiring the existence of large scale $M$, associated with new physics.
Indeed, since $v\approx 175$ GeV, $m_\nu \approx 0.03$ eV, for $M\approx
10^{15}$ GeV. Thus Neutrino masses are a probe of physics at grand
unification mass scale. We shall see that neutrino oscillations might
remarkably provide a mechanism to measure extremely small masses (of order
of milli electron volts and less) and indirectly provide a new scale
indicative of new physics.

\section{Neutrino Oscillations}

\subsection{Oscillations in vacuum}

Neutrinos are produced in weak interactions as flavor eigenstates,
characterized by $e,\mu ,\tau $. The flavor eigenstates $\left| \nu _\alpha
\right\rangle $ need not coincide with mass (energy) eigenstate $\left| \nu
_i\right\rangle $ and are generally coherent superposition of such states 
\begin{equation}
\left| \nu _\ell \right\rangle =\sum_iU_{\ell i}\left| \nu _i\right\rangle
\label{e17}
\end{equation}
where the mixing matrix is unitary. This matrix is characterized by 3
angles, $\theta _{12}=\theta _3,\theta _{13}=\theta _2,\theta _{23}=\theta
_1 $, one CP violating phase $\delta $ and two Majorana phases, which we put
equal to zero. 
\begin{equation}
U=\left( 
\begin{tabular}{ccc}
$c_{12}c_{13}$ & $s_{12}c_{13}$ & $s_{13}e^{-i\delta }$ \\ 
$-s_{12}c_{23}-c_{12}s_{23}s_{13}e^{i\delta }$ & $%
c_{12}c_{23}-s_{12}s_{23}s_{13}e^{i\delta }$ & $s_{23}c_{13}$ \\ 
$s_{12}s_{23}-c_{12}c_{23}s_{13}e^{i\delta }$ & $%
-s_{23}c_{12}-s_{12}c_{23}s_{13}e^{i\delta }$ & $c_{23}c_{13}$%
\end{tabular}
\right)  \label{cU}
\end{equation}
In vacuum, the mass eigenstates propagate as plane waves 
\begin{equation}
\left| \nu _i\left( t,{\bf x}\right) \right\rangle =\exp \left( -i\left(
E_it-{\bf k}\cdot {\bf x}\right) \right) \left| \nu _i\left( 0\right)
\right\rangle  \label{e19}
\end{equation}
where $E_i\approx k+m_i^2/2k^2$, $k\gg m_i$. Thus flavor eigenstates
propagate as 
\begin{equation}
\left| \nu _\ell \left( t,{\bf x}\right) \right\rangle =U_{\ell i}\exp
\left( -i\left( E_it-{\bf k}\cdot {\bf x}\right) \right) \left| \nu _i\left(
0\right) \right\rangle  \label{e20}
\end{equation}
The probability at time $t$ that $\nu _\ell $ is converted into $\nu _{\ell
^{\prime }}$ is 
\[
P_{\nu _\ell \rightarrow \nu _{\ell ^{\prime }}}=\left| \left\langle \nu
_{\ell ^{\prime }}|\nu _\ell \right\rangle \right| ^2 
\]
For oscillations involving two neutrinos, it takes a simple form 
\begin{eqnarray}
P_{\nu _\ell \rightarrow \nu _{\ell ^{\prime }}} &=&\left| \left\langle \nu
_{\ell ^{\prime }}|\nu _\ell \right\rangle \right| ^2  \nonumber \\
&=&\sin ^2\theta \cos ^2\theta \left| 1+\exp \left[ -\left( E_1-E_2\right)
t\right] \right| ^2  \nonumber \\
&=&\sin ^22\theta \sin ^2\left[ \left( \frac{E_1-E_2}2\right) t\right] 
\nonumber \\
&=&\sin ^22\theta \sin ^2\left[ \frac{\Delta m^2}{4k}t\right]  \label{e21}
\end{eqnarray}
It is convenient to write it as 
\begin{equation}
P_{\nu _\ell \rightarrow \nu _{\ell ^{\prime }}}=\sin ^22\theta \sin
^2\left[ 1.27\frac{\Delta m^2}{E_\nu }L\right]  \label{e22}
\end{equation}
where $L$ is the distance (measured in meters) travelled after $\nu _\ell $
is converted into $\nu _{\ell ^{\prime }}$. $\Delta m^2=m_1^2-m_2^2$ in
units of eV$^2$ while $E_\nu \simeq k$ is measured in MeV. Thus the
oscillations in this simple case are characterized by the oscillation length
in vacuum 
\begin{equation}
L_v=4\pi \frac{E_\nu }{\Delta m^2}  \label{e23}
\end{equation}
and by the amplitude $\sin ^22\theta $. To look for the oscillations, the
above formula also shows that one needs low energy $\nu $'s, long path
length and large flux. In the flavor basis the Hamiltonian is 
\begin{equation}
H_\nu =UHU^{-1}  \label{e24}
\end{equation}
where $H$ is diagonal in $\nu _1-\nu _2$ base. This gives, ignoring a
trivial diagonal term not relevent for oscillations, 
\begin{equation}
H_\nu =2\pi \left( 
\begin{tabular}{cc}
$-\frac{\cos 2\theta }{L_v}$ & $\frac{\sin 2\theta }{2L_v}$ \\ 
$\frac{\sin 2\theta }{2L_v}$ & $0$%
\end{tabular}
\right)  \label{e25}
\end{equation}

\subsection{Oscillations in Matter}

In traversing matter neutrinos interact with electrons and nucleons of
intervening material and their forward coherent scattering induces an
effective potential energy $\sqrt{2}G_FN_e$ modifying $H_\nu $ given in Eq. (%
\ref{e21}) to 
\begin{equation}
H_M=\left( 
\begin{tabular}{cc}
$-\frac{\cos 2\theta }{L_v}+\frac 1{L_0}$ & $\frac{\sin 2\theta }{2L_v}$ \\ 
$\frac{\sin 2\theta }{2L_v}$ & $0$%
\end{tabular}
\right)  \label{e27}
\end{equation}
Thus the evolution of the flavor eigenstates in matter is governed by the
Schr\"{o}dinger equation $\left[ x=ct=t\right] $%
\begin{equation}
i\frac d{dx}\left( 
\begin{array}{c}
\nu _e \\ 
\nu _\alpha
\end{array}
\right) =2\pi H_M\left( 
\begin{array}{c}
\nu _e \\ 
\nu _\alpha
\end{array}
\right)  \label{e26}
\end{equation}
where $H_M$ is given in Eq. (\ref{e27}) and there 
\begin{equation}
L_0=\frac{2\pi }{\sqrt{2}G_FN_e}=1.7\times 10^7\left( m\right) /\rho \left(
g/cm^3\right) Y_e  \label{e28}
\end{equation}
is the corresponding matter oscillation length. Here $N_e$ denotes the
number of electrons per unit volume: 
\begin{equation}
N_e=\frac \rho {m_N}Y_e  \label{e29}
\end{equation}
where $Y_e$ is the number of electrons per nucleons $\approx 1/2$ in
ordinary matter. The effective oscillation length in matter is 
\begin{eqnarray}
L_m &=&L_v\frac{\sin 2\theta _m}{\sin 2\theta }=L_v\left[ \left( \frac{L_v}{%
L_0}-\cos 2\theta \right) ^2+\sin ^22\theta \right] ^{-1/2}  \nonumber \\
\tan 2\theta _m &=&\tan 2\theta \left( 1-\frac{L_v}{L_0\cos 2\theta }\right)
^{-1}  \nonumber \\
P_{\left( \nu _e\rightarrow \nu _e\right) } &=&1-\sin ^22\theta _m\sin
^2\left[ 1.27\frac L{L_m}\right]  \label{e30}
\end{eqnarray}
$\theta _m$ is new mixing angle in matter. Thus, resonance $\left[ \sin
^22\theta _m=1\right] $ occures when $\cos 2\theta $ is equal to 
\begin{equation}
\frac{L_v}{L_0}=\frac{2\sqrt{2}G_FN_eE_\nu }{\Delta m^2}=0.22\left[ \frac{%
E_\nu }{1\,MeV}\right] \left[ \frac{\rho Y}{100g/cm^3}\right] \left[ \frac{%
7\times 10^{-5}eV^2}{\Delta m^2}\right]  \label{e31}
\end{equation}
The transition point between the regime of vacuum and matter oscillations is
determined by the ratio $L_v/L_0$. If it is greater than 1, matter
oscillations dominate. If it is less than $\cos 2\theta $, vacuum
oscillation dominate Generally there is a smooth transition between these
two regimes. Matter effects become maximum at resonance $L_v/L_0=\cos
2\theta $. This is the basis of Mikheyev-Smirnov-Wolfenstein (MSW) effect.
The survival probability $P_{\left( \nu _e\rightarrow \nu _e\right) }$
averaged over the detector position $L$ (from the solar surface) \cite{r06}
is 
\begin{equation}
P_{\left( \nu _e\rightarrow \nu _e\right) }=\frac 12\left[ 1+\left(
1-2P_x\right) \cos 2\theta _m\left( \rho _{\max }\right) \cos 2\theta \right]
\label{e32}
\end{equation}
where $\theta _m\left( \rho _{\max }\right) $ is the initial mixing angle,
usually 
\[
\cos 2\theta _m\left( \rho _{\max }\right) \simeq -1 
\]
and $P_x$ is a finite probability for jumping from one eigenstate to the
other one and convertion might be incomplete. The survival probability $%
\left\langle P_{\left( \nu _e\rightarrow \nu _e\right) }\right\rangle $ as a
function of $E_\nu $ is displayed for various mixing angles in Fig. 3 \cite
{r07}.

For the parameters corresponding to prefered solution for neutrino
oscillations (see below) $\sin ^22\theta \approx 0.8$, $\Delta m^2\approx
7\times 10^{-5}$ eV$^2$ and $\rho =100$ g/cm$^3$ at the center of the sun, $%
Y_e\approx 1/2$, $L_v/L_0=0.11E_v/1$ MeV, $\cos 2\theta =0.45$.

Due to different reaction thersholds, solar neutrinos with energy $E_\nu
>0.814$ MeV can be detected in $^{37}Cl$ and those with $E_\nu >0.233$ MeV
in $^{71}Ga$. Note that for pp neutrinos ($E_\nu <0.$42 MeV) and $^7Be$
neutrinos ($E_\nu \approx 0.$86 MeV), $L_v/L_0<\cos 2\theta $ and they
undergo vacuum oscillations, while the neutrinos with $E_\nu >4.5$ MeV, ($%
^8B $ neutrinos) undergo MSW matter oscillations.

\section{Evidence for Oscillations}

One looks for oscillations in two types of experiments.

\subsection{Appearance experiments:}

Here one searches for a new neutrino flavor, absent in the initial beam,
which can arise from oscillations.

\subsubsection{Atmospheric neutrino anomaly:}

Atmospheric neutrinos are produced in decays of pions (kaons) that are
produced in the interaction of cosmic rays with the atmosphere: 
\begin{eqnarray*}
p+A &\rightarrow &\pi ^{\pm }+A^{\prime } \\
\pi ^{\pm } &\rightarrow &\mu ^{\pm }\nu _\mu \left( \bar{\nu}_\mu \right) \\
&\rightarrow &e^{\pm }\nu _e\left( \bar{\nu}_e\right) \nu _\mu \left( \bar{%
\nu}_\mu \right)
\end{eqnarray*}
These neutinos are detected in and beneath underground detectors through the
reactions 
\begin{eqnarray*}
\nu _\mu +n &\rightarrow &\mu ^{-}+p \\
\bar{\nu}_\mu +p &\rightarrow &\mu ^{+}+n
\end{eqnarray*}
and 
\begin{eqnarray*}
\nu _e+n &\rightarrow &e^{-}+p \\
\bar{\nu}_e+p &\rightarrow &e^{+}+n
\end{eqnarray*}
These are respectively called $\mu $-like and $e$-like events. The observed
ratios of these events was found to be substantially reduced from the
expected value $\sim 2$. There is compelling evidence that atmospheric
neutrinos change flavor as the Super-Kamiokande experiment clearly indicated
a deficit of up-ward $\mu $-like events (produced about 10$^4$ km away at
the opposite side of earth) relative to the downward going events (produced
about 20 km above). The $e$-like events showed a normal zenith angle
dependence. The data is described by $\nu _\mu \rightarrow \nu _\tau $
oscillations. The conversion probability $P_{\nu _\mu \rightarrow \nu _\tau
} $ fits the data quite well for \cite{r08} 
\begin{equation}
\Delta m_{23}^2=2.0\times 10^{-3}eV^2,\,\,\sin ^22\theta _{23}\approx 1.0
\label{e33}
\end{equation}

\subsubsection{Solar Neutrinos}

Particularly compelling evidence that the solar neutrinos change flavor has
been reported by the Sudbury Neutrino Observatory (SNO). SNO measures the
high energy part of the solar neutrino flux ($^8B$ neutrinos). The reactions 
\begin{eqnarray*}
\nu d &\rightarrow &\nu np \\
&\rightarrow &epp \\
\nu e &\rightarrow &\nu e
\end{eqnarray*}
were studied by SNO. SNO measured arriving $\nu _e+\nu _\mu +\nu _\tau $
flux, $\phi _e+\phi _{\mu \tau }$, and the $\nu _e$ flux, $\phi _e$. From
the observed rates for the first two reactions, which involve respectively
neutral current and charge current, SNO finds that the ratio of the two
fluxes $\phi _e$ and $\phi _e+\phi _{\mu \tau }$ is $0.306\pm \ 0.050$. This
implies that the flux $\phi _{\mu \tau }$ is not zero. Since all the
neutrinos are born in nuclear reactions that produce only electron
neutrinos, it is clear that neutrinos change flavor. Corroborating
information comes from the direct reaction $\nu e\rightarrow \nu e$, studied
by both SNO and Super-Kamiokande. The strongly favored explanation of solar
neutrino flavor change is the Large Mixing Angle version of the MSW effect,
with the best fit parameters \cite{r09} 
\begin{equation}
\Delta m_{12}^2=7.1\times 10^{-5}eV^2,\,\,\,\,\sin ^22\theta
_{12}=0.8\,\,\left[ {See Fig. 3}\right]  \label{e34}
\end{equation}

\subsection{Disappearance experiments:}

Reactors are source of $\bar{\nu}_e$'s through the neutron $\beta $-decay 
\[
n\rightarrow p+e^{-}+\bar{\nu}_e 
\]
and experiment looks for a possible decrease in the $\bar{\nu}_e$ flux as a
function of distance from the reactor, $\bar{\nu}_e\rightarrow X$ [if
converted to $\bar{\nu}_\mu $, say, one would see nothing, $\bar{\nu}_\mu $
could have produced $\mu ^{+}$ but does not have sufficient energy to do
so]. Kamland experiment \cite{r10} confirms that $\bar{\nu}_e$ do indeed
disappear when the reactor $\bar{\nu}_e$ have travelled $\approx 200$ km. $%
\bar{\nu}_e$ flux is only $0.611\pm \ 0.085\pm \ 0.041$ of what it would be
if none of it were disappearing. Interestingly this reactor $\bar{\nu}_e$
disappearance and the solar neutrino results can be described by the same
neutrino mass and mixing parameters (see Fig. 4 \cite{r07}). This gives
confidence that the physics of both phenomenon has been correctly identified.

\section{Neutrino Mass Matrix}

As discussed the data from solar and atmospheric neutrino and reactor
antineutrinos experiments provide evidence for neutrino mass and mixing with
two different mass scales and large mixing angles: 
\begin{eqnarray}
\Delta m_{atm}^2 &\equiv &\Delta m_{23}^2=\left( 2.0\pm 0.5\right) \times
10^{-3}eV^2  \nonumber \\
\sin ^2\theta _{23} &\equiv &\sin ^2\theta _1=1.00\pm 0.4  \nonumber \\
\Delta m_{solar}^2 &\equiv &\Delta m_{12}^2=\left( 7.1\pm 0.6\right) \times
10^{-5}eV^2  \nonumber \\
\tan ^2\theta _{12} &\equiv &\tan ^2\theta _3=0.45\pm 0.06  \label{e35}
\end{eqnarray}
Further the CHOOZ experiment \cite{r11} gives 
\begin{equation}
\left| U_{e3}\right| ^2\equiv \sin ^2\theta _2<4\times 10^{-2}  \label{e36}
\end{equation}

We would interpret these results in terms of small off-diagonal
perturbations of a degenerate diagonal mass matrix in flavor basis for light
Majorana neutrinos \cite{r12}. In this approach there is no fundamental
distinction between masses of neutrinos of different flavors; the mass
differences arise from small flavor violation of off-diagonal Yukawa
coupling constants. Further the neutrino mass differences do not in anyway
constraint the absolute value of neutrino mass. The constraint on it will
come from neutrinoless double $\beta $-decay experiments, cosmology and
direct laboratory experiments, e.g. tritium $\beta $-decay.

Let us consider a Majorana mass matrix in $\left( e,\mu ,\tau \right) $
basis 
\begin{equation}
m_\nu =m_0\left( 
\begin{tabular}{ccc}
$a_{ee}$ & $a_{e\mu }$ & $a_{e\tau }$ \\ 
$a_{e\mu }$ & $a_{\mu \mu }$ & $a_{\mu \tau }$ \\ 
$a_{e\tau }$ & $a_{\mu \tau }$ & $a_{\tau \tau }$%
\end{tabular}
\right)  \label{e37}
\end{equation}
It is convenient to define the neutrino mixing angles as follows 
\begin{equation}
\left( 
\begin{array}{c}
\nu _e \\ 
\nu _\mu \\ 
\nu _\tau
\end{array}
\right) =U\left( 
\begin{array}{c}
\nu _1 \\ 
\nu _2 \\ 
\nu _3
\end{array}
\right)  \label{e38}
\end{equation}
where $U$ is given in Eq. (\ref{cU}). We shall put $\delta $ as well as
Majorana phases to be zero. In view of mixng angles given above, we shall
take $s_{13}\equiv s_2=0$, $c_{13}\equiv c_2=1$. and $c_1=1/\sqrt{2}$, $%
s_1=\mp 1/\sqrt{2}$. The diagonalization gives 
\begin{eqnarray}
a_{e\mu } &=&\pm a_{e\tau }=\frac 1{\sqrt{2}}s_3c_3\left( -m_1+m_2\right) 
\nonumber \\
a_{\mu \tau } &=&\pm \frac 12\left[ \left( m_1s_3^2+m_2c_3^2\right)
-m_3\right]  \nonumber \\
a_{\mu \mu } &=&a_{\tau \tau }=\frac 12\left[ m_1s_3^2+m_2c_3^2+m_3\right] 
\nonumber \\
a_{ee} &=&m_1s_3^2+m_2c_3^2.  \label{e39}
\end{eqnarray}
In view of 
\begin{equation}
\Delta m_{12}^2=m_2^2-m_1^2\ll \Delta m_{23}^2=m_3^2-m_2^2,  \label{e40}
\end{equation}
we can take 
\[
m_1\simeq \pm m_2. 
\]
Thus we have two possiblities for mass matrix $m_1=m_2=m_0;$ $m_1=-m_2=m_0$: 
\begin{eqnarray}
m_\nu &=&m_0\left( 
\begin{tabular}{ccc}
$1$ & $0$ & $0$ \\ 
$0$ & $\frac 12\left( 1+a\right) $ & $\pm \frac 12\left( 1-a\right) $ \\ 
$0$ & $\pm \frac 12\left( 1-a\right) $ & $\frac 12\left( 1+a\right) $%
\end{tabular}
\right)  \label{e41} \\
m_\nu &=&m_0\left( 
\begin{tabular}{ccc}
$\cos 2\theta _3$ & $-\frac 1{\sqrt{2}}\sin 2\theta _3$ & $\mp \frac 1{\sqrt{%
2}}\sin 2\theta _3$ \\ 
$-\frac 1{\sqrt{2}}\sin 2\theta _3$ & $\frac 12\left( \cos 2\theta
_3+a\right) $ & $\pm \frac 12\left( \cos 2\theta _3-a\right) $ \\ 
$\mp \frac 1{\sqrt{2}}\sin 2\theta _3$ & $\pm \frac 12\left( \cos 2\theta
_3-a\right) $ & $\frac 12\left( \cos 2\theta _3+a\right) $%
\end{tabular}
\right)  \label{e42}
\end{eqnarray}
where $a=m_3/m_0$. If we do not want to commit to any particular value of $%
\theta _3$, then we have the first case with the following subcases
corresponding to $m_0=0$, $a=-1,1,-2,2,0$ respectively \cite{r13} 
\[
\begin{tabular}{ccc}
$i)$ &  & $\frac{m_3}2\left( 
\begin{tabular}{ccc}
$0$ & $0$ & $0$ \\ 
$0$ & $1$ & $1$ \\ 
$0$ & $1$ & $1$%
\end{tabular}
\right) $ \\ 
$ii)$ &  & $m_0\left( 
\begin{tabular}{ccc}
$1$ & $0$ & $0$ \\ 
$0$ & $0$ & $1$ \\ 
$0$ & $1$ & $0$%
\end{tabular}
\right) $ \\ 
$iii)$ &  & $m_0\left( 
\begin{tabular}{ccc}
$1$ & $0$ & $0$ \\ 
$0$ & $1$ & $0$ \\ 
$0$ & $0$ & $1$%
\end{tabular}
\right) $ \\ 
$iv)$ &  & $\frac{m_0}2\left( 
\begin{tabular}{ccc}
$2$ & $0$ & $0$ \\ 
$0$ & $-1$ & $3$ \\ 
$0$ & $3$ & $-1$%
\end{tabular}
\right) $ \\ 
$v)$ &  & $\frac{m_0}2\left( 
\begin{tabular}{ccc}
$2$ & $0$ & $0$ \\ 
$0$ & $3$ & $-1$ \\ 
$0$ & $-1$ & $3$%
\end{tabular}
\right) $ \\ 
$vi)$ &  & $\frac{m_0}2\left( 
\begin{tabular}{ccc}
$2$ & $0$ & $0$ \\ 
$0$ & $1$ & $1$ \\ 
$0$ & $1$ & $1$%
\end{tabular}
\right) $
\end{tabular}
\]
In order to generate $\Delta m_{12}^2$ and $\Delta m_{23}^2$, we will now
concentrate on choice (iii), which preserves flavor and add to it a small
perturbation which violates flavor in off-diagonal matrix elements: 
\begin{equation}
m_\nu =m_0\left( 
\begin{tabular}{ccc}
$1$ & $\varepsilon _{12}$ & $\varepsilon _{13}$ \\ 
$\varepsilon _{12}$ & $1$ & $\varepsilon _{23}$ \\ 
$\varepsilon _{13}$ & $\varepsilon _{23}$ & $1$%
\end{tabular}
\right)  \label{e43}
\end{equation}
where $\varepsilon _{ij}\ll 1$. The diagonalization gives 
\begin{equation}
m_i=m_0\left( 1-x_i\right)  \label{e44}
\end{equation}
where $x_i\left( i=1,2,3\right) $ are roots of cubic equation 
\begin{equation}
x^3-\left( \varepsilon _{12}^2+\varepsilon _{13}^2+\varepsilon
_{23}^2\right) x+2\varepsilon _{12}\varepsilon _{13}\varepsilon _{23}=0.
\label{e45}
\end{equation}
The choice $\varepsilon _{12}=\varepsilon _{13}=\varepsilon
_{23}=\varepsilon $ will give the roots $\left( \varepsilon ,\varepsilon
,-2\varepsilon \right) $ and thus will not lift the degeneracy between $m_1$
and $m_2$. To lift this degeneracy we take $\varepsilon _{12}=\varepsilon
_{13}=\varepsilon +\delta $, $\varepsilon _{23}=\varepsilon $ with $\delta
/\varepsilon \ll 1$. Then the roots to the first order in $\delta
/\varepsilon $ are $\varepsilon \left( 1+\frac 43\frac \delta \varepsilon
\right) $, $\varepsilon $, $-2\varepsilon \left( 1+\frac 23\frac \delta
\varepsilon \right) $ so that 
\begin{eqnarray}
m_1 &=&m_0\left[ 1-\varepsilon -\frac 43\delta \right]  \nonumber \\
m_2 &=&m_0\left[ 1-\varepsilon \right]  \nonumber \\
m_3 &=&m_0\left[ 1+2\varepsilon +\frac 43\delta \right]  \nonumber \\
\Delta m_{12}^2 &\approx &\frac 83m_0^2\delta \left( 1-\varepsilon \right)
\simeq \frac 83m_0^2\delta  \nonumber \\
\Delta m_{23}^2 &\approx &6\varepsilon m_0^2.  \label{e46}
\end{eqnarray}
This gives 
\begin{eqnarray}
\frac \delta \varepsilon &=&\frac 94\frac{\Delta m_{12}^2}{\Delta m_{23}^2}%
\simeq 5.9\times 10^{-2}  \nonumber \\
\sqrt{\varepsilon }m_0 &\simeq &2.1\times 10^{-2}{ eV}  \label{e47}
\end{eqnarray}

Thus $m_0$ is not constrained. However, $m_0$ is constrained by WMAP data, $%
3m_0<0.71$ eV. When analyzed in conjunction with neutrino oscillation, it is
found that mass eigenvalues are essentially degenerate with $3m_0>0.4$ eV.
The above limits put limits on $\varepsilon $: $7.9\times
10^{-3}<\varepsilon <2.5\times 10^{-2}$.

For the degenerate neutrino mass pattern $m_1\sim m_2\sim m_3\gg \sqrt{%
\Delta m_{32}^2}=0.045$, the effective mass in neutrinoless double $\beta $%
-decay is larger than $\sim 0.05$ eV, constrained from above by the mass
limit from tritium $\beta $-decay. If the effective Majorana mass is
confirmed to be $\left( 0.39_{-0.28}^{+0.17}\right) $ eV \cite{r02}, it
would strongly indicate that neutrinos follow degenrate mass pattern (see
Fig. 5 \cite{r07}), when 
\[
\frac{\Delta m^2}{m^2}\ll 1. 
\]

Finally for two modest extensions of the standard model in which the
neutrino mass matrix advocated in this section can be embedded, see Ref. 
\cite{r12}.

\section{Conclusion}

To conclude various neutrino mass patterns and corresponding neutrino mass
matrix types are possible. Further the absolute value of neutrino mass is
not yet determined. However, one thing is certain that neutrinos are
providing an evidence for new physics but the scale of new physics is not
yet pinned down. The heavy right handed neutrinos at new physics scale may
provide an explanation for baryogenesis through leptogenesis. If past is of
any guide, neutrinos will enrich physics still further.

Acknowledgements: The author would like to thank the warm hospitality
extended to him at MTPR-04 by Dr. Lotfia Alnadi and Professor M. M. Sherif
which enabled the author to enjoy his atay in Egypt -- a unique place
comprising Ancient, Islamic and Modern Civilizations. This work was
supported in part by a grant from Pakistan Council for Science and
Technology.

\section{Figure Captions}

\begin{enumerate}
\item  The mass spectrum of quarks and leptons we do not understand.

\item  Composition of the Universe.

\item  Schematic illustration of the survival probability of $\nu _e$
created at the solar center. The curves are labelled by the $\sin ^22\theta $
values.

\item  Ratio of observed to expected rates (without neutrino oscillations)
for reactor neutrino experiments as a function of distance, including the
recent result from the Kamland experiment. The shaded region is that
expected due to neutrino oscillations with large mixing parameters as
determined from solar neutrino data.

\item  Dependence of effective Majorana mass $\left\langle m_\nu
\right\rangle $ derived from the rate of neutrinoless double $\beta $-decay
on the absolute mass of the lightest neutrino. The stripes region indicates
the range related to the unknown Majorana phases, while the cross hatched
region is covered if one $\sigma $ errors on the oscillation parameters also
included. The arrows indicate the three possible neutrino mass patterns.
\end{enumerate}

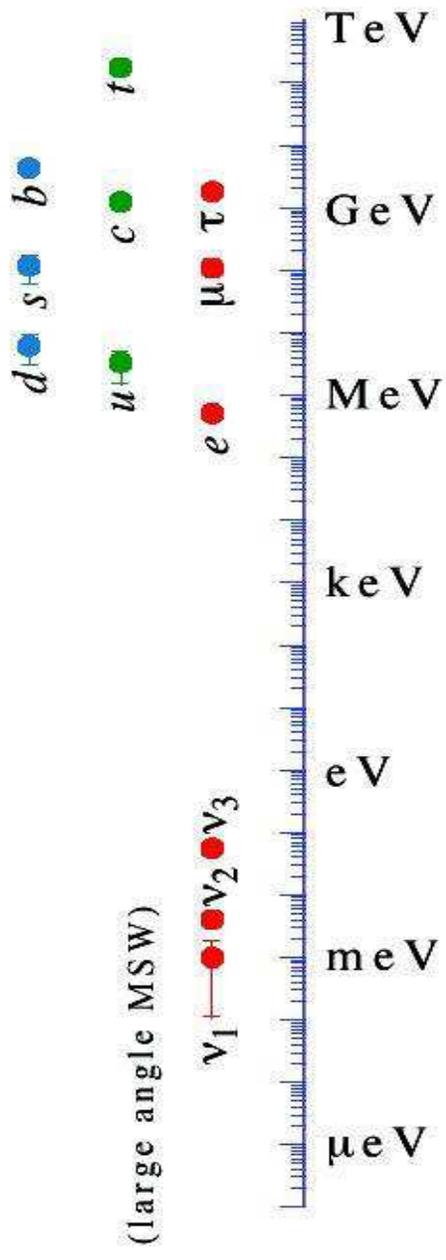
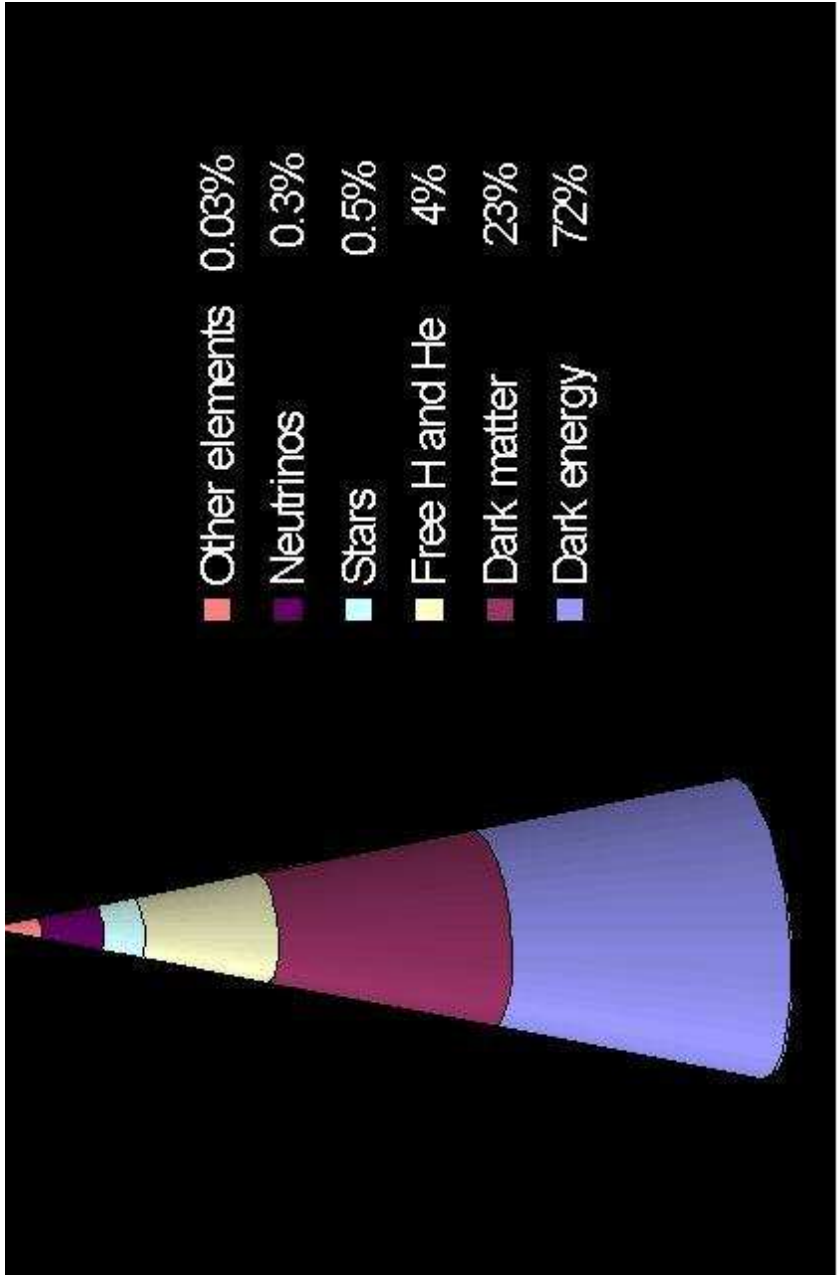
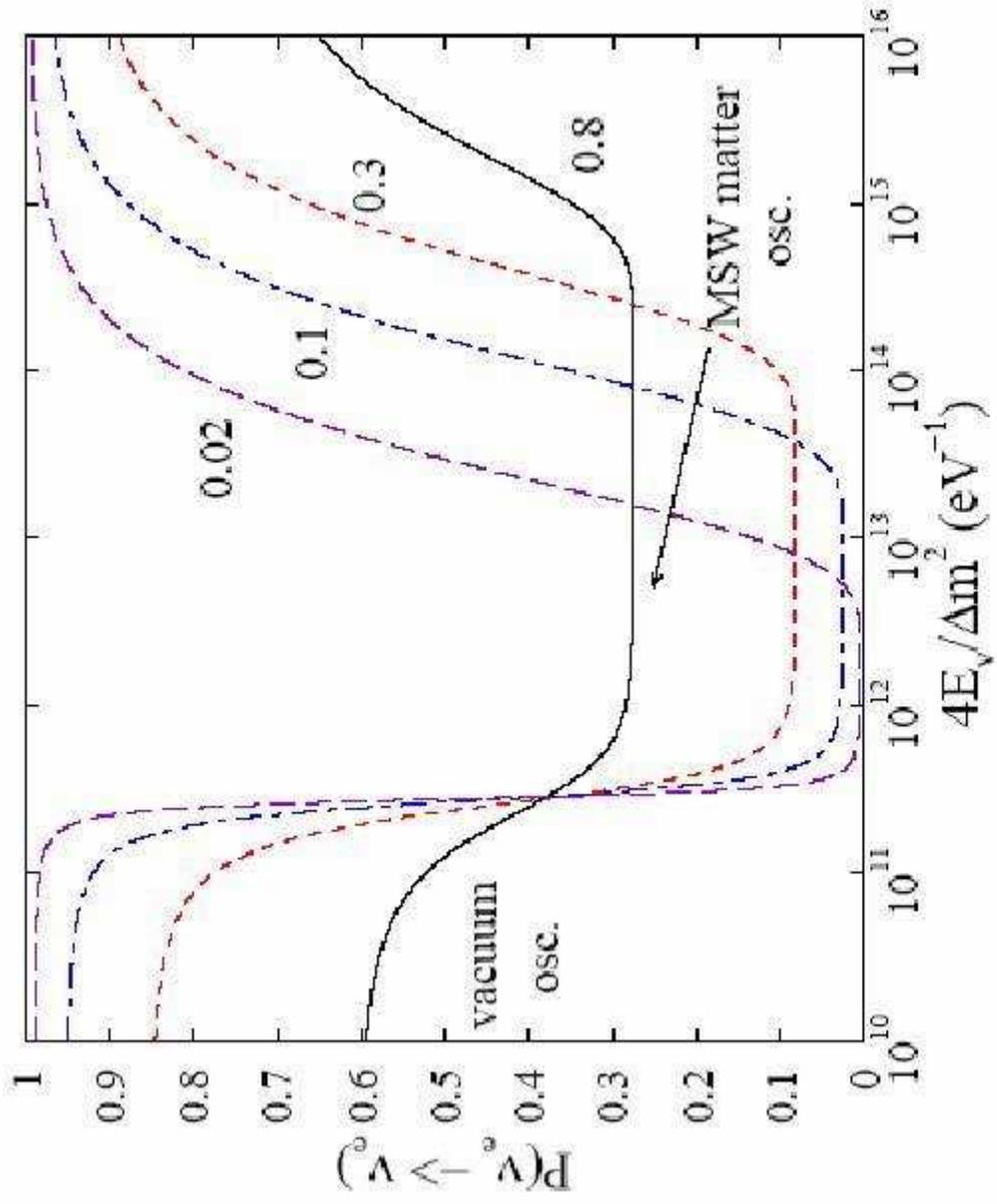
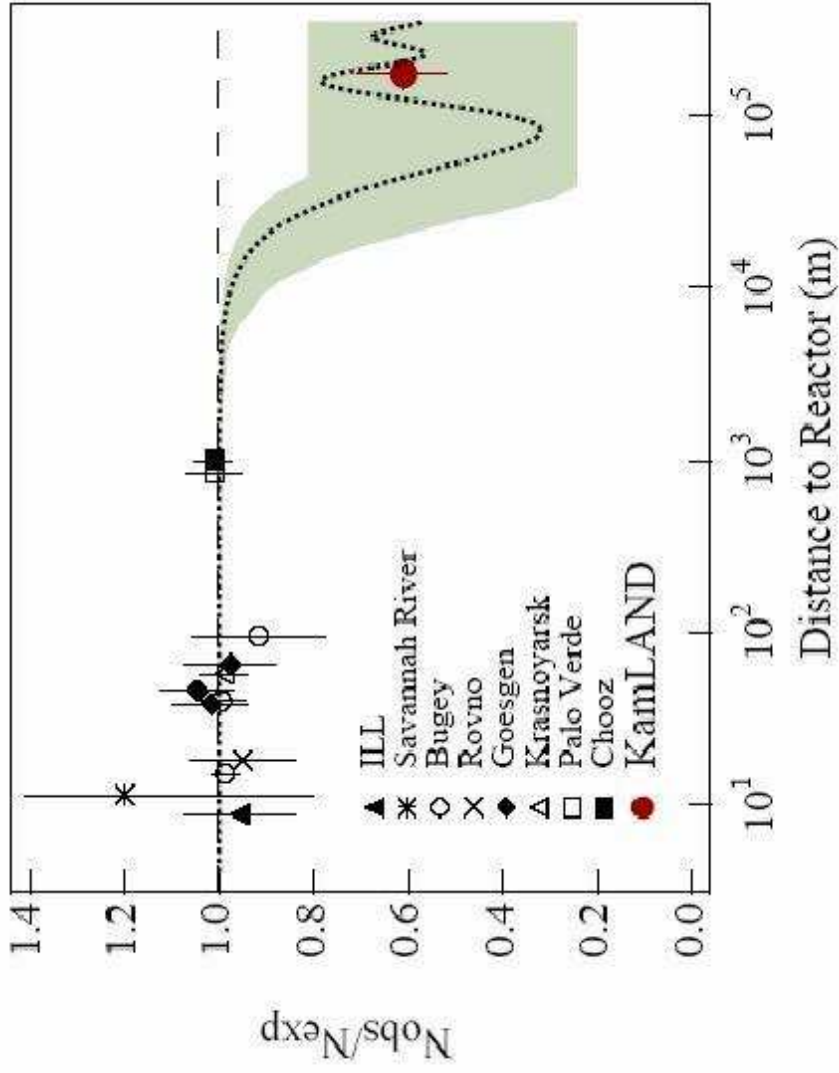
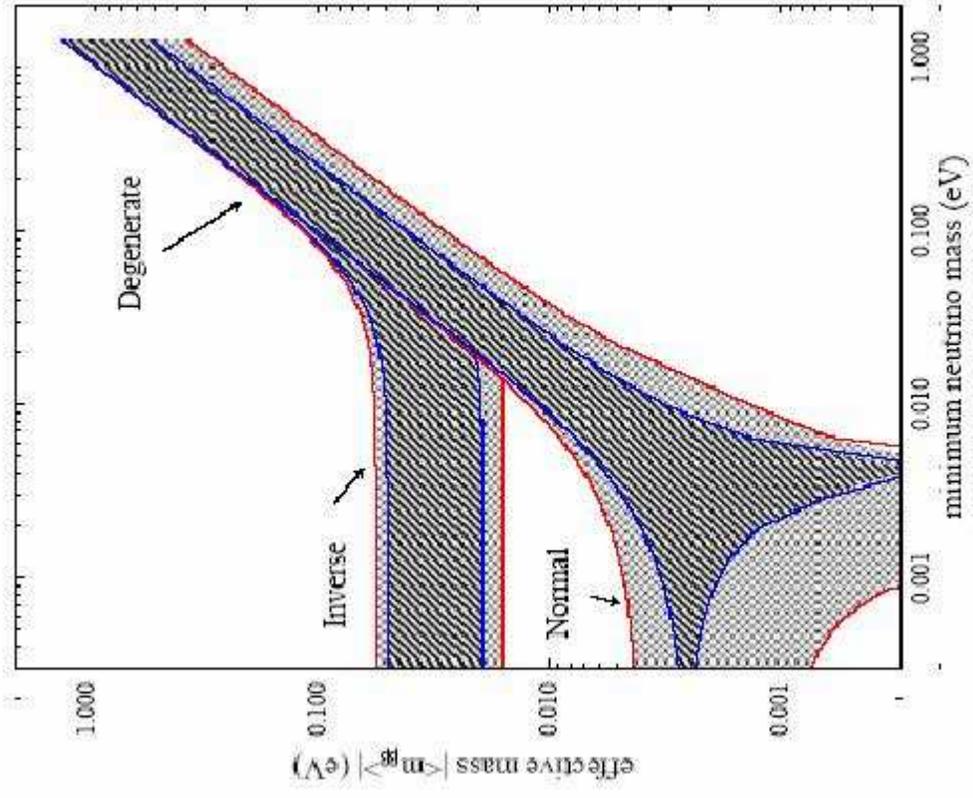
\end{document}